# A particular case of a Lifting Hele-Shaw Cell

I.V. Grossu[1], S. A. El Shamali[2]

[1] University of Bucharest, Faculty of Physics, Bucharest-Magurele, Romania
[2] University of Agronomical Sciences and Veterinary Medicine, Faculty of Horticulture, Bucharest, Romania

We present a very simple method of obtaining spectacular fractals, using a particular case of a lifting Hele-Shaw cell.

In this paper we simply describe a particular case of a lifting Hele-Shaw cell [1-4]. The device is very easy to assemble and the result is spectacular.

We need only two windowpanes with the same size. On each one, we must disperse a film of paint, water (tempera, acrylic) or even oil paint. The mixture must be enough dense, but also fluid, too much or to less solvent will not conduce to success. The combination of colors will contribute to the artistic effect.

In a second stage, we must put together the two films, pressing firmly the windowpanes. After the sheets are well stocked, we must separate them by introducing, for example, a knife between the glasses. In case of failure, we can retry, by repeating the last steps (the second stage). Finally, two symmetrical fractals [5], one on each side, must be obtained (Fig.1, 2).

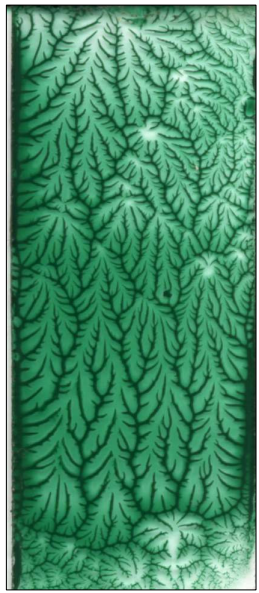 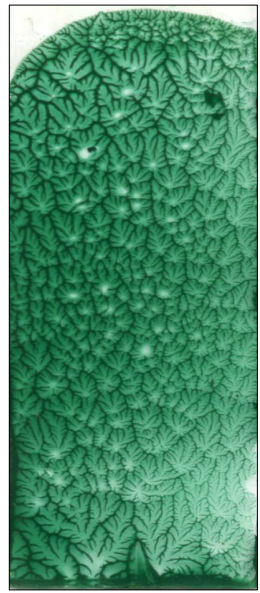

a  b

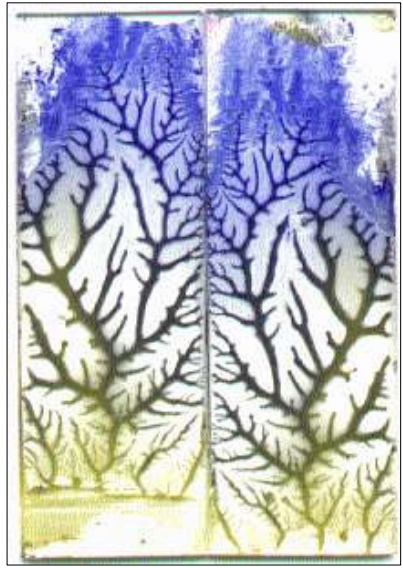
c

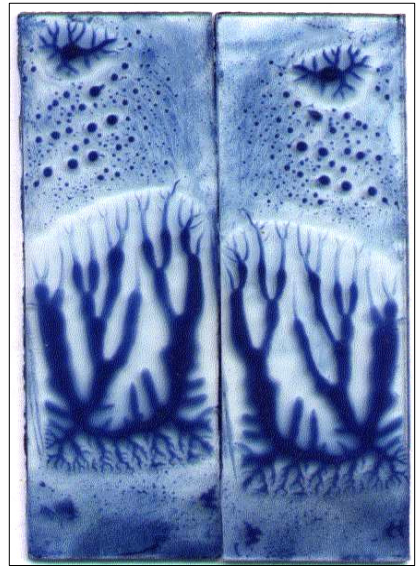
d

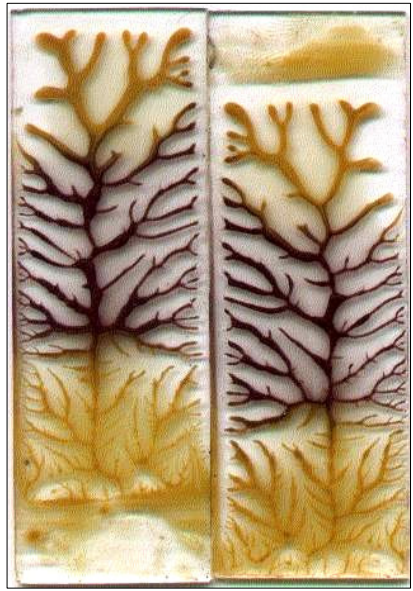
e

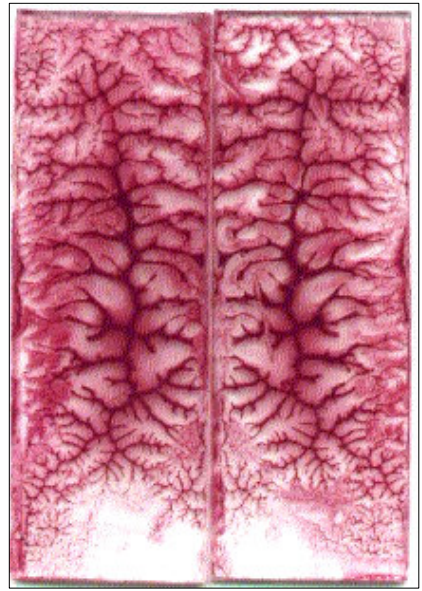
f

Fig.1. Water paints (a-e); oil paint (f).

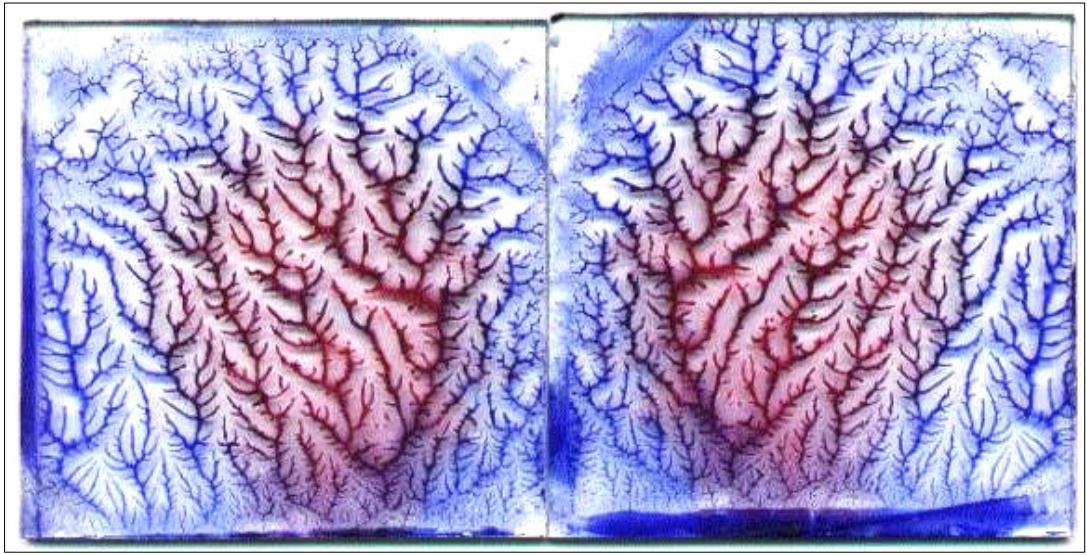

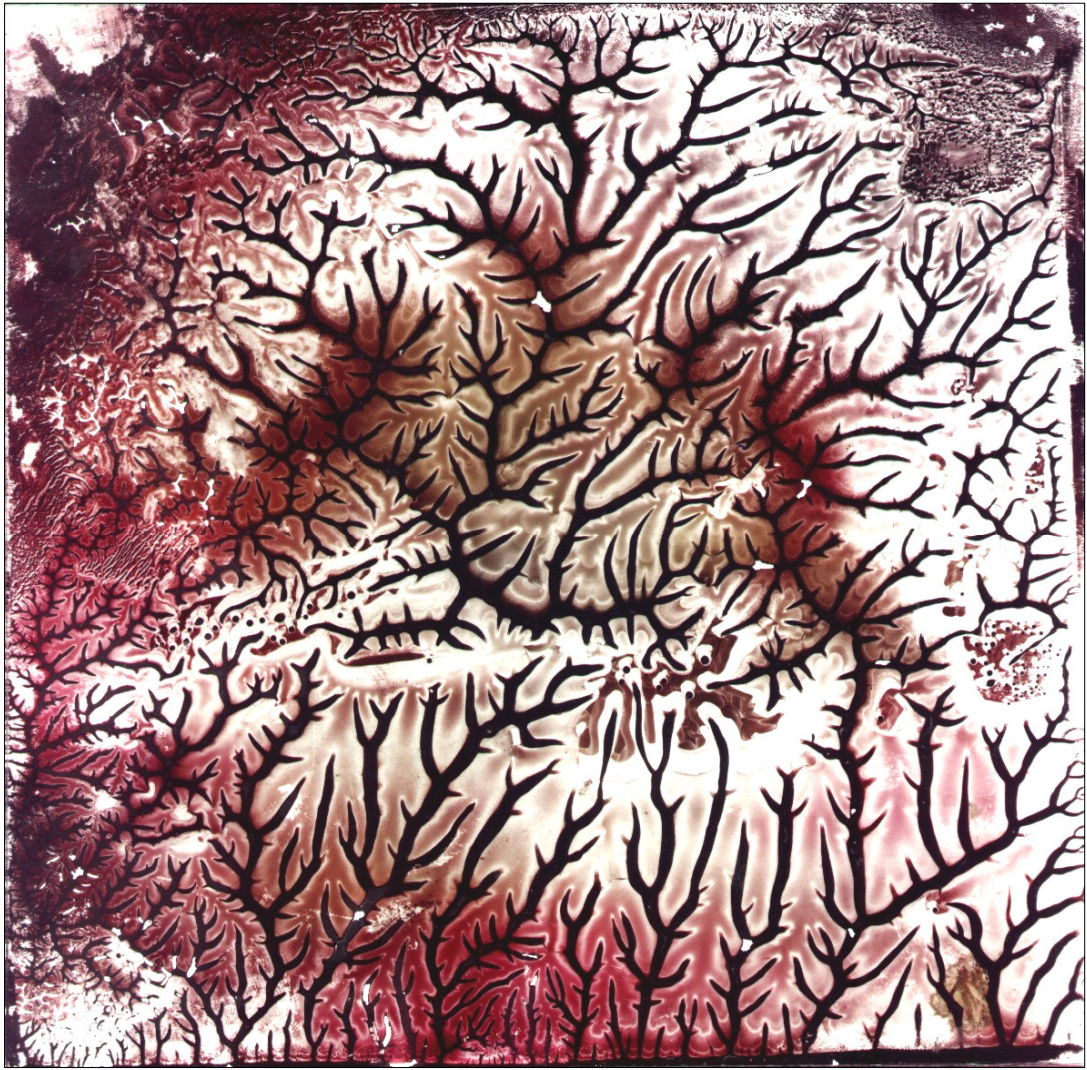

Fig.2. The effect was tested using windowpanes with different sizes, up to 50cm.